\documentclass[12pt]{article}

\usepackage[dvips]{graphicx}
\usepackage{amsmath,amssymb}

\begin{document}
\baselineskip=15pt

\def\unt#1{\,{\rm #1}}
\def\tref#1#2#3{{\bf #1} (#2) #3}
\def\bref#1{(\ref{#1})}
\newcommand{\ii}{\ensuremath{\mathrm{i}}}
\newcommand{\e}{\ensuremath{\mathrm{e}}}
\newcommand{\ppi}{\ensuremath{\mathrm{\pi}}}
\newcommand{\dif}[1]{\ensuremath{ \mathrm{d}#1 }}
\newcommand{\borel}{\ensuremath{\mathcal{B}}}
\newcommand{\il}{\ensuremath{\mathcal{L}^{-1}}}
\newcommand{\qcd}{\ensuremath{\mathrm{QCD}}}

\begin{center}

{\Huge \bf --------------------------}

{\tiny 

SUNY Institute of Technology at Utica/Rome\\

Conference on Theoretical High Energy Physics\\

June 6th, 2002
}


{\Huge \bf --------------------------}

\end{center}
\vskip 1cm

\begin{center}
{\Large\bf
Gaussian Sum-Rules, Scalar Gluonium, and Instantons
}

\vskip .8cm

T.G.\  {\sc Steele $^{a,}$}
\footnote{Electronic address: {\tt Tom.Steele@usask.ca}}
,
D.\ {\sc Harnett $^{a}$},
\\ 
G.\ {\sc Orlandini $^{b,}$}

\vskip .8cm
{\small \it
$^a$ 
Department of Physics \& Engineering Physics, University of Saskatchewan, 
Saskatoon, Saskatchewan, S7N 5E2, Canada
\\
$^b$ 
Dipartimento di Fisica and INFN Gruppo Collegato di Trento,
Universit\`a di Trento, I-38050 Povo, Italy
}
\end{center}

\vskip 1cm	

\begin{center}
{\Large \bf ------------------------------------------------------------}
\end{center}
\vskip 1cm	

\begin{center}
{\bf  Abstract}\\

\end{center}

{\small \sl

Gaussian sum-rules relate a QCD prediction to a two-parameter Gaussian-weighted integral of a hadronic spectral function, providing
 a clear conceptual connection to quark-hadron duality.  
In contrast to Laplace sum-rules, the Gaussian sum-rules 
exhibit enhanced  sensitivity to excited 
 states of the hadronic spectral function.
The formulation of Gaussian sum-rules and associated analysis techniques 
for extracting hadronic properties from the sum-rules are reviewed and applied to 
scalar gluonium. With the inclusion of instanton effects, the Gaussian sum-rule analysis results in a consistent scenario
where the gluonic resonance strength is spread over  a broad energy range below $1.6\unt{GeV}$, and indicates the 
presence of  gluonium content in more than one 
hadronic state.

}

\sf

\newpage

\setcounter{footnote}{0}
\setcounter{equation}{0}
\setcounter{section}{0}

\section{Introduction}\label{intro_sec}
The hadronic spectrum has too many scalar states above $1\,{\rm GeV}$ for a $q\bar q$ nonet, as would be anticipated if  gluonium states
exist  in the 1--2 GeV region \cite{pdg}.  Determining how this gluonium content is distributed among these scalar-isoscalar resonances is thus an important issue. 
In particular, the possibility that the observed hadronic states are mixtures of gluonium and quark mesons must be explored.

Gaussian sum-rules  are sensitive to the hadronic spectral function over a wide energy range, and are thus well-suited to studying the distribution of 
gluonium states. The simplest Gaussian sum-rule (GSR) \cite{gauss}
\begin{equation}
G\left( \hat s,\tau\right)=\frac{1}{\ppi}\int\limits_{t_0}^\infty\frac{1}{\sqrt{4\ppi\tau}}
\exp{\left( -\frac{\left(t-\hat s\right)^2}{4\tau}\right)}\,\rho(t)\,\mathrm{d}t
\quad ,\quad \tau>0
\label{basic_gauss}
\end{equation}
relates a QCD prediction on the left-hand side of \bref{basic_gauss} to the hadronic spectral function $\rho(t)$ (with physical threshold $t_0$)
smeared over the energy range $\hat s -2\sqrt{\tau}\lesssim t\lesssim\hat s +2\sqrt{\tau}$, representing an energy interval for quark-hadron duality.  An interesting 
aspect of the GSR is that the duality interval is constrained by QCD. A lower bound on this duality scale $\tau$ necessarily exists because the 
QCD prediction has 
renormalization-group properties that reference running quantities the the energy scale $\nu^2=\sqrt{\tau}$ \cite{gauss,orl00}.
Thus it is not possible to achieve the formal  $\tau\to 0$ limit where complete knowledge of the spectral function could be obtained  via
\begin{equation}
\lim_{\tau\to 0}G\left(\hat s,\tau\right)=\frac{1}{\ppi}\rho\left(\hat s\right)\quad ,\quad \hat s>t_0 \quad .
\label{gauss_limit}
\end{equation}
However, there is no theoretical constraint on the quantity $\hat s$ representing the peak of the Gaussian kernel appearing in \bref{basic_gauss}.  Thus the
$\hat s$ dependence of the QCD prediction $G\left(\hat s,\tau\right)$ probes the behaviour of the smeared spectral function, reproducing the essential 
features of the spectral function. In particular, as $\hat s$ passes through $t$ values corresponding to  resonance peaks, the Gaussian kernel in 
\bref{basic_gauss} reaches its maximum value, implying that Gaussian sum-rules weight excited and ground states equally.  
This behaviour should be contrasted with Laplace sum-rules 
\begin{equation}
R\left(\Delta^2\right)=\frac{1}{\pi}\int\limits_{t_0}^\infty\exp{\left(-\frac{t}{\Delta^2}\right)}\rho(t)\, dt
\quad ,
\label{basic_laplace}
\end{equation}
where excited states are damped by the exponential decay of the Laplace kernel.  Thus, in comparison with Laplace sum-rules the Gaussian sum-rule (GSR) 
has an enhanced sensitivity  to excited states of the spectral function.

In this paper, the original formulation of Gaussian sum-rules  \cite{gauss}  and analysis techniques for extracting spectral function hadronic features 
\cite{orl00,har01}  will be reviewed.  These techniques will then be applied to scalar gluonium, where instanton contributions are known to be 
crucial for a consistent Laplace sum-rule analysis \cite{shuryak,gluelet}.  Results of the GSR analysis indicate that the gluonium spectral strength
is distributed across a broad energy range below $1.6\,{\rm GeV}$ \cite{har01}.

\section{Foundations of Gaussian Sum-Rules}\label{gauss_sec}
The general formulation  of GSRs will be reviewed in the context of scalar gluonium probed by the following correlation function.
\begin{gather}
  \Pi(Q^2)  = \ii\int\mathrm{d}^4\!x\, \e^{\ii q\cdot x} \langle O | T\{J(x),J(0)\} | O\rangle
  \quad,\quad Q^2=-q^2 
\label{corr}\\
  J(x) = -\frac{\ppi^2}{\alpha\beta_0} \beta(\alpha)  G^a_{\ \mu\nu}(x)  G^{a\mu\nu}(x)\quad ,\quad
 G^a_{\ \mu\nu} = \partial_{\mu} A^a_{\nu} - \partial_{\nu} A^a_{\mu}
  + g f^{abc} A^b_{\mu} A^c_{\nu}
\label{current}\\
  \beta\left(\alpha\right) =\nu^2\frac{\mathrm{d}}{\mathrm{d}\nu^2}\left(\frac{\alpha(\nu)}{\ppi}\right)=
  -\beta_0\left(\frac{\alpha}{\ppi}\right)^2-\beta_1\left(\frac{\alpha}{\ppi}\right)^3
  +\ldots
  \label{beta_expansion}
\\
  \beta_0=\frac{11}{4}-\frac{1}{6} n_f\quad ,\quad
  \beta_1=\frac{51}{8}-\frac{19}{24}n_f   \quad,\quad\ldots
  \label{beta_coefficients}
\end{gather}
The current $J(x)$ is renormalization-group invariant in the chiral limit of $n_f$ massless quarks as needed to probe physical 
(renormalization-group invariant) hadronic states.

\begin{figure}[htb]
\centering
\includegraphics[scale=0.3,angle=270]{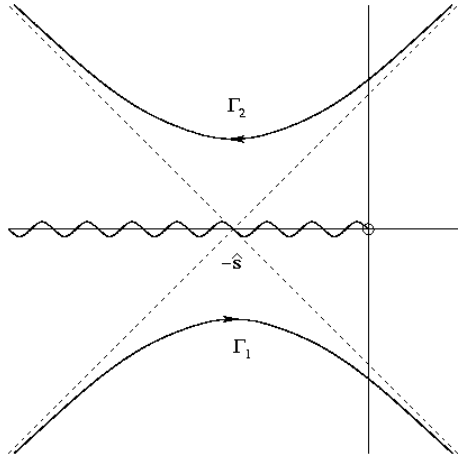}
\caption{{\small Contour of integration $\Gamma_1+\Gamma_2$ defining the
Gaussian  sum-rule.
The wavy line
on the negative real axis denotes the branch cut of $\Pi(z)$.
}}
\label{cont_fig}
\end{figure}

From the asymptotic form and assumed analytic properties of~(\ref{corr})
follows a dispersion relation with three subtraction constants
\begin{equation}\label{dispersion}
  \Pi(Q^2) - \Pi(0) =   Q^2\Pi'(0) + \frac{1}{2}Q^4 \Pi''(0)
     -\frac{Q^6}{\ppi} \int_{t_0}^{\infty} \frac{\rho(t)}{t^3 (t+Q^2)} \dif{t} \quad ,\quad Q^2 >0
\end{equation}
where $\rho(t)$ is the hadronic spectral function  with physical threshold $t_0$, and the subtraction constant $\Pi(0)$ has  been included on the side of the equation containing the QCD prediction
because it is determined  by the low-energy theorem~\cite{let}
\begin{equation}\label{let}
  \Pi(0)\equiv\lim_{Q^2\rightarrow 0} \Pi(Q^2) = \frac{8\pi}{\beta_0} \langle J\rangle\quad.
\end{equation}
The undetermined subtraction constants $\Pi'(0)$ and $\Pi''(0)$ and field-theoretical divergences in $\Pi(Q^2)$  
proportional to $Q^4$ are eliminated  in an integer-weighted family of
Gaussian sum-rules\footnote{This definition is a natural generalization of that given in~\cite{gauss}.
To recover the original Gaussian sum-rule, we simply let $k=0$ in~(\ref{srdef}).}
\begin{equation}\label{srdef}
   G_k(\hat{s},\tau)\equiv \sqrt{\frac{\tau}{\ppi}}\borel
   \left\{ \frac{(\hat{s}+\ii\Delta)^k \Pi(-\hat{s}-\ii\Delta)
         - (\hat{s}-\ii\Delta)^k \Pi(-\hat{s}+\ii\Delta) }{\ii\Delta}
   \right\} 
\end{equation}
where $ k= -1,0,1, \ldots$ and  with the Borel transform $\borel$ defined by
\begin{equation}\label{borel}
  \borel\equiv \lim_{\stackrel{N,\Delta^2\rightarrow\infty}{\Delta^2/N\equiv 4\tau}}
  \frac{(-\Delta^2)^N}{\Gamma(N)}\left( \frac{\mathrm{d}}{\mathrm{d}\Delta^2}\right)^N \quad.
\end{equation}
Applying definition~(\ref{srdef})  to both sides of~(\ref{dispersion})
 annihilates the undetermined low-energy constants
and the field theoretical divergence contained in $\Pi(Q^2)$.
 Using the identity
\begin{equation}\label{identity}
  \borel\left[ \frac{(\Delta^2)^n}{\Delta^2 +a}\right] =
  \frac{1}{4\tau} (-a)^n \exp\left( \frac{-a}{4\tau}\right)\ \mathrm{for}\ n\geq 0  \quad,
\end{equation}
then leads to the following GSR family:
\begin{equation}
 G_{k}(\hat{s},\tau)+ \delta_{k\,-1}\frac{1}{\sqrt{4\pi\tau}} \exp\left( \frac{-\hat{s}^2}{4\tau}\right)
   \Pi(0) 
= \int_{t_0}^{\infty} t^k
    \exp\left[ \frac{-(\hat{s}-t)^2}{4\tau}\right] \frac{1}{\pi} \rho(t) \dif{t}  
\label{gauss_family}
\end{equation}
Calculation of the Borel transform is achieved through an identity relating~(\ref{borel}) to the inverse Laplace transform~\cite{gauss}
\begin{equation}\label{bor_to_lap}
  \borel [f(\Delta^2)] = \frac{1}{4\tau}\il [f(\Delta^2)]
\end{equation}
where, in our notation,
\begin{equation}
  \il [f(\Delta^2)] = \frac{1}{2\ppi \ii} \int_{a-\ii\infty}^{a+\ii\infty}
    f(\Delta^2) \exp\left( \frac{\Delta^2}{4\tau} \right) \dif{\Delta^2}
\end{equation}
with $a$ chosen such that all singularities of $f$ lie to the left of $a$
in the complex $\Delta^2$-plane.
With a change of variables, the calculation of the GSR reduces to \cite{har01}
\begin{equation}
  G_k(\hat{s},\tau)  = \frac{1}{\sqrt{4\ppi\tau}}\frac{1}{2\ppi \ii}
  \int_{\Gamma_1 +\Gamma_2} (-w)^k  \exp
  \left[ \frac{-(\hat{s}+w)^2}{4\tau}\right] \Pi(w) \dif{w}
  \label{finish}
\end{equation}
where $\Gamma_1$ and $\Gamma_2$ are the parabolae depicted in Figure~\ref{cont_fig}.

\section{Normalized Gaussian Sum-Rules}
Studies of Gaussian sum-rules have traditionally focussed on their connection with finite-energy sum-rules as   established  through the diffusion
equation
\begin{equation}
\frac{\partial^2 G_k\left( \hat s,\tau\right)}{\partial\hat s^2}=
\frac{\partial G_k\left( \hat s,\tau\right)}{\partial \tau}\quad .
\label{diffusion}
\end{equation}
In particular, the resonance(s) plus continuum model
\begin{equation}
  \rho(t)=\rho^{{\rm had}}(t)+\theta\left(t-s_0\right){\rm Im}\Pi^{\qcd}(t)\quad ,
\label{respcont}
\end{equation}
when $\rho^{{\rm had}}(t)$ is evolved through the diffusion equation \bref{diffusion}, only reproduces the QCD prediction at large energies ($\tau$ large)
if the resonance and continuum contributions are balanced through the finite-energy sum-rules \cite{gauss}
\begin{equation}
F_n\left(s_0\right)=\frac{1}{\pi}\int\limits_{t_0}^{s_0} t^n\rho^{\rm had}(t)\,\mathrm{d}t \quad ,\quad n={\rm integer}  \quad .
\label{basic_fesr}
\end{equation}

Within the resonance(s) plus continuum model \bref{respcont}, the continuum contribution to the GSRs is determined by QCD
\begin{equation}\label{continuum}
   G_k^{{\rm cont}} (\hat{s},\tau,s_0) = \frac{1}{\sqrt{4\pi\tau}}   \int_{s_0}^{\infty} t^k
   \exp \left[ \frac{-(\hat{s}-t)^2}{4\tau} \right]  \frac{1}{\pi} {\rm Im} \Pi^{\qcd}(t) \dif{t}\quad ,
\end{equation}
and is thus combined with $G_k\left(\hat s,\tau\right)$ to give the total QCD contribution
\begin{equation}
  G_k^{\qcd}\left(\hat{s},\tau,s_0\right) \equiv G_k\left(\hat{s},\tau\right) -  G_k^{{\rm cont}} \left(\hat{s},\tau,s_0\right) \quad ,
\label{blah} 
\end{equation}
resulting in the final relation between the QCD and hadronic sides of the GSRs.
\begin{equation}
 G_{k}^\qcd\left(\hat{s},\tau,s_0\right)+ \delta_{k\,-1}\frac{1}{\sqrt{4\pi\tau}} \exp\left( \frac{-\hat{s}^2}{4\tau}\right)
   \Pi(0) 
= \int_{t_0}^{\infty} t^k
    \exp\left[ \frac{-(\hat{s}-t)^2}{4\tau}\right] \frac{1}{\pi} \rho^{\rm had}(t) \dif{t}  
\label{final_gauss}
\end{equation}

Integrating both sides of \bref{final_gauss} reveals that the normalization of the GSRs is related to the finite-energy sum-rules
\begin{equation}
  \int\limits_{-\infty}^\infty G_k^{\qcd}(\hat{s}, \tau,s_0) \dif{\hat{s}}+\delta_{k\,-1}\Pi(0)
  =\frac{1}{\pi}\int\limits_{t_0}^{\infty} t^k\rho^{{\rm had}}(t) \dif{t} \quad .
  \label{tom_norm_2}
\end{equation}
Thus the diffusion equation analysis \cite{gauss} relates the normalization of the GSR to the finite-energy sum-rules.  Information independent of this 
relation is extracted from the normalized GSRs
\begin{gather}
  N^{\qcd}_k (\hat{s}, \tau, s_0) \equiv
  \frac{G^{\qcd}_{k} (\hat s, \tau, s_0) + \delta_{k\,-1}\frac{1}{\sqrt{4\pi\tau}} \exp\left( \frac{-\hat{s}^2}{4\tau}\right)
   \Pi(0) }{M^{\qcd}_{k,0} (\tau, s_0)+\delta_{k\,-1}\Pi(0)}
  \label{tom_norm_srk}\\
  M_{k,n}(\tau, s_0)
  = \int\limits_{-\infty}^\infty \hat{s}^n G_k (\hat s,\tau, s_0) \dif{\hat{s}}
  \quad,\quad n=0,1,2,\ldots\quad,
\label{moments}
\end{gather}
which are related to the hadronic spectral function via  
\begin{equation}\label{ngsr}
   N_k^{\qcd}(\hat{s},\tau,s_0) = \frac{ \frac{1}{\sqrt{4\pi\tau}} \int_{t_0}^{\infty} t^k
   \exp\left[\frac{-(\hat{s}-t)^2}{4\tau} \right] \frac{1}{\pi}\rho^{{\rm had}}(t)\dif{t}}{\int_{t_0}^{\infty} t^k
   \frac{1}{\pi}\rho^{{\rm had}}(t)} \quad.
\end{equation}

\section{Gaussian Sum-Rule Analysis Techniques}
In the single narrow resonance model, $\rho^{{\rm had}}(t)$ takes the form
\begin{equation}\label{single}
  \rho^{{\rm had}}(t) = \pi f^2 \delta(t-m^2)
\end{equation}
where $m$ and $f$ are respectively the resonance mass and coupling. With such an ansatz, the
normalized Gaussian sum-rule~(\ref{ngsr}) becomes
\begin{equation}\label{phenom_single}
  N_k^{\qcd}(\hat{s},\tau,s_0) = \frac{1}{\sqrt{4\ppi\tau}}
  \exp\left[ -\frac{(\hat{s}-m^2)^2}{4\tau}\right]
  \quad.
\end{equation}
Deviations from the narrow-width limit are  proportional to $m^2\Gamma^2/\tau$, so this narrow-width model may actually be a good numerical approximation.  Phenomenological analysis of the single narrow resonance model proceeds from the observation that \bref{phenom_single} has a maximum value (peak) 
at $\hat s=m^2$ independent of the value of $\tau$.  The value of $s_0$ is then optimized by minimizing the $\tau$ dependence of the $\hat s$ 
peak position of the QCD prediction, and the resulting $\tau$-averaged $\hat s$ peak position leads to a prediction of the resonance mass \cite{orl00}.

The $\rho$ meson illustrates this single-resonance analysis technique and demonstrates that GSRs can be used to predict resonance properties.  The correlation function of the vector-isovector correlation function results in the following ($k=0$) GSR
\begin{equation}
\begin{split}
G_0^{\qcd}\left(\hat s, \tau, s_0\right)
=&\frac{1}{16\pi^2} \left(1+\frac{\alpha\left(\sqrt{\tau}\right)}{\pi}\right)
\left[{\rm erf}\left(\frac{\hat s}{2\sqrt{\tau}}\right)
+{\rm erf}\left(\frac{s_0-\hat s}{2\sqrt{\tau}}\right)
\right]
\\
&-\frac{\hat s}{32\pi^2\tau\sqrt{\pi\tau}}
{\exp{\left(-\frac{\hat s^2}{4\tau}\right)}}
\left\langle C^v_4{\cal O}^v_4\right\rangle
\\
&+\frac{1}{64\pi^2\tau\sqrt{\pi\tau}}\left(-1+\frac{\hat s^2}{2\tau}\right)
{\exp{\left(-\frac{\hat s^2}{4\tau}\right)}}
\left\langle C_6^v{\cal O}^v_6\right\rangle
\\
&-\frac{\hat s}{128\pi^2\tau^2\sqrt{\pi\tau}}\left(-1+\frac{\hat s^2}{6\tau}\right)
{\exp{\left(-\frac{\hat s^2}{4\tau}\right)}}
\left\langle C_8^v{\cal O}^v_8\right\rangle\quad ,
\end{split}
\nonumber
\label{gauss_qcd_vec}
\end{equation}     
where 
\begin{gather}
\left\langle C^v_4{\cal O}^v_4\right\rangle=\frac{\pi}{3}\left\langle \alpha G^2\right\rangle
-8\pi^2 m\left\langle \bar q q\right\rangle
\label{c4_vec}\\
\left\langle C^v_6{\cal O}^v_6\right\rangle=-\frac{896}{81}\pi^3\alpha\left(\left\langle
\bar q q\right\rangle\right)^2
\label{c6_vec}
\end{gather}
and  $SU(2)$ symmetry along with the vacuum saturation hypothesis have been employed.  For brevity, we refer 
to the literature \cite{dim_eight} for the 
expressions 
for the dimension eight condensates, and simply use \bref{gauss_qcd_vec} to establish a convention
consistent with \cite{bordes}.  Note that the running coupling is referenced to the energy scale $\sqrt{\tau}$, a point which will be discussed in the next Section.

The non-perturbative QCD condensate contributions in \bref{gauss_qcd_vec} are exponentially suppressed for large
$\hat s$.  Since $\hat s$ represents the location of the Gaussian peak on the phenomenological side of the sum-rule,
the non-perturbative corrections are most important in the low-energy region, as anticipated by the 
role of  QCD condensates in relation to the vacuum properties of QCD.  This explicit low-energy role of the QCD
condensates clearly exhibited for the Gaussian sum-rules is obscured in the Laplace sum-rules.  

\begin{figure}[htb]
\centering
\includegraphics[scale=0.5,angle=270]{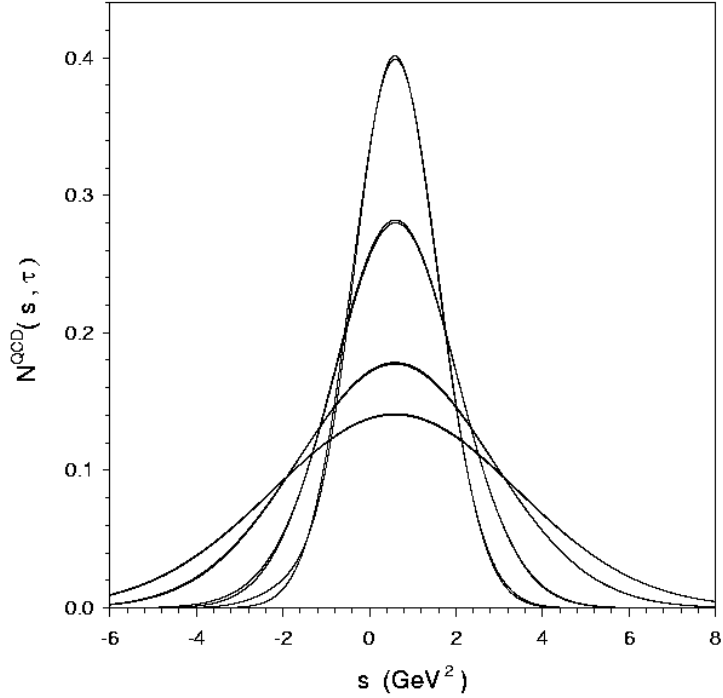}
\caption{{ \small Comparison of the vector-current 
theoretical prediction for  $N_0^{{\rm QCD}}\left(\hat s, \tau,s_0\right)$ with the
single resonance phenomenological model.
The $\tau$ values used for the four pairs of curves, from top to bottom in the figure, are respectively
$\tau=0.5\unt{GeV^4}$, $\tau=1.0\unt{GeV^4}$, $\tau=2.5\unt{GeV^4}$, and $\tau=4.0\unt{GeV^4}$. Note the almost complete overlap between the QCD 
prediction and phenomenological model. }
}
\label{rho_fig}
\end{figure}

The QCD inputs used for the $\rho$ meson analysis analysis  are  
\begin{gather}
\Lambda^{(3)}=300\,{\rm MeV}
\\
\left\langle \alpha G^2\right\rangle=\left(0.045\pm 0.014\right)\unt{GeV^4}\quad ,
\quad 2m\left\langle \bar q q\right\rangle
=-f_\pi^2m_\pi^2
\\
\left\langle C_6^v{\cal O}^v_6\right\rangle=-f_{vs}\frac{896}{81}\pi^3\left(1.8\times 10^{-4}\unt{GeV^6}\right)
,~ f_{vs}=1.5\pm 0.5
\\
\left\langle C_8^v{\cal O}^v_8\right\rangle=\left(0.40\pm 0.16\right)\unt{GeV^8}
\quad,
\end{gather}
consistent with the condensate parameters in \cite{bordes}.  The
 criteria of $\tau$ stability of $\hat s$ peak  predicts  the following values for $m_\rho$  and $s_0$ \cite{orl00}
\begin{equation}
m_\rho=(0.75\pm 0.07)\unt{GeV}\quad ,\quad s_0=(1.2\pm 0.2)\unt{GeV^2}
\end{equation}
in excellent agreement with the known value of the $\rho$ mass.
Furthermore, the 
 phenomenological and QCD sides of the (normalized) Gaussian sum-rules shown in  Figure \ref{rho_fig} are in superb agreement even for very small values 
of $\tau$ \cite{orl00}. This agreement is particularly impressive since there are no free parameters corresponding to the normalization of the curves in this analysis.

In more complicated resonance models, the $\hat s$ peak position of the phenomenological model begins to develop $\tau$ dependence which is well-described by \cite{orl00,har01}
\begin{equation}
\hat s_{peak}(\tau,s_0)=  A + \frac{B}{\tau} + \frac{C}{\tau^2}\quad .
\label{peak_drift}
\end{equation}
Analysis of how the $\hat s$ peak ``drifts'' with $\tau$ in comparison with the behaviour \bref{peak_drift} then allows optimization  of $s_0$.
After optimization of $s_0$,  the resonance model parameters are extracted from various moments of the GSRs.  For example, 
a square pulse\footnote{This model would describe a broad, structureless  feature in the hadronic spectral function.}  
centred at $t=m^2$ with total width $2m\Gamma$ leads to the following ($k=0$) normalized GSR \cite{har01}
 \begin{equation}
 N^{\qcd}_0(\hat{s},\tau,s_0)
=\frac{1}{4m\Gamma}\left[
{\rm erf}\left(\frac{\hat s-m^2+m\Gamma}{2\sqrt{\tau}}\right)
-{\rm erf}\left(\frac{\hat s-m^2-m\Gamma}{2\sqrt{\tau}}
\right)
\right]\quad .
\label{gauss_sp}
\end{equation}
Expansion of \bref{gauss_sp} for small $\Gamma$ demonstrates that deviations from the narrow width limit \bref{phenom_single} scale as $m^2\Gamma^2/\tau$.
The resonance parameters can then be determined by the following  moment combinations [see \bref{moments}] of the right-hand side of \bref{gauss_sp} \cite{har01} 
\begin{gather}
  \frac{M_{0,1}}{M_{0,0}} = m^2 
\label{square_params_a}
\\
  \sigma_0^2-2\tau   \equiv \frac{M_{0,2}}{M_{0,0}} -\left(\frac{M_{0,1}}{M_{0,0}}\right)^2 ={\frac{1}{3}m^2 \Gamma^2} \quad,
\label{square_params_b}
\end{gather}
where it is understood that the QCD expressions at the optimized value of $s_0$ are used on the left-hand side.\footnote{The residual moment combinations leading to the resonance parameters must be $\tau$ independent, providing a consistency check on the analysis.  Weak residual $\tau$ dependence is averaged over the $\tau$ range used to extract the optimized $s_0$ from the peak-drift analysis.}  

Moments also provide a method for testing the accuracy of agreement between the QCD and phenomenological sides of the normalized GSR beyond a simple $\chi^2$ measure which could be extracted from plots such as Figure \ref{rho_fig}.  For example, a combination of third-order moments representing the asymmetry 
of the $\hat s$ dependence about its average value results in \cite{har01}
\begin{equation}
  A_0^{(3)} =
\frac{M_{0,3}}{M_{0,0}} - 3\left(\frac{M_{0,2}}{M_{0,0}}\right)
  \left( \frac{M_{0,1}}{M_{0,0}}\right) + 2\left(\frac{M_{0,1}}{M_{0,0}}\right)^3
= 0 \quad,
\end{equation}
and hence a deviation of the QCD value of this moment from its value of  $A_0^{(3)} =0$  in the square pulse model indicates a shortcoming of the model spectral function in comparison with QCD.  

The procedure for studying an $N$-parameter resonance model is easily generalized. The peak-drift analysis is used to optimize $s_0$,  
the lowest $N$  moments are used to determine the resonance model 
parameters, and the next-highest moment combination is employed as a test of the accuracy of the model's agreement with the QCD prediction.

\section{Scalar Gluonium Gaussian Sum-Rules}
The lowest two Gaussian sum-rules for scalar gluonium contain perturbative, condensate and instanton corrections,    
and are given by \cite{har01}
\begin{equation}\label{GM1}
\begin{split}
     G_{-1}^{\qcd}(\hat{s},\tau,s_0) = &-\frac{1}{\sqrt{4\pi\tau}} \int_0^{s_0} t\, \dif{t}
     \exp\left[ \frac{-(\hat{s}-t)^2}{4\tau}\right]  \Biggl[ (a_0-\pi^2 a_2)
     +2a_1\log\left( \frac{t}{\nu^2} \right)\Biggr.
\\
&\phantom{-\frac{1}{\sqrt{4\pi\tau}} \int_0^{s_0} t
     \exp\left[ \frac{-(\hat{s}-t)^2}{4\tau}\right]} \qquad \Biggl.+ 3a_2 \log^2\left( \frac{t}{\nu^2} \right) \Biggr]
   \\
      &+ \frac{1}{\sqrt{4\pi\tau}} \exp\left( \frac{-\hat{s}^2}{4\tau}\right)
       \left[ -b_0 \langle J\rangle + \frac{c_0 \hat{s}}{2\tau}\left\langle {\cal O}_6\right\rangle
      -\frac{d_0}{4\tau}\left(\frac{\hat{s}^2}{2\tau} -1 \right)
      \left\langle {\cal O}_8\right\rangle \right]
   \\
      &-\frac{16\pi^3}{\sqrt{4\pi\tau}} \int\dif{n}(\rho)\rho^4
      \int_0^{s_0} t  \exp\left[ \frac{-(\hat{s}-t)^2}{4\tau}\right]  J_2\left(\rho\sqrt{t} \right)
      Y_2\left(\rho\sqrt{t} \right)  \dif{t}
   \\
      &-\frac{128\pi^2}{\sqrt{4\pi\tau}} \exp\left( \frac{-\hat{s}^2}{4\tau}\right)
      \int\dif{n}(\rho)
\end{split}
\end{equation}
\begin{equation}\label{G0}
\begin{split}
      G_0^{\qcd}(\hat{s},\tau,s_0) = &- \frac{1}{\sqrt{4\pi\tau}} \int_0^{s_0} t^2 \dif{t}
       \exp\left[ \frac{-(\hat{s}-t)^2}{4\tau}\right]  \Biggl[ (a_0-\pi^2 a_2)
       +2a_1\log\left( \frac{t}{\nu^2} \right)\Biggr. 
\\ 
&\phantom{- \frac{1}{\sqrt{4\pi\tau}} \int_0^{s_0} t^2 \dif{t}
       \exp\left[ \frac{-(\hat{s}-t)^2}{4\tau}\right]}\qquad \Biggl.+ 3a_2 \log^2\left( \frac{t}{\nu^2} \right) \Biggr]      
   \\
       &- \frac{1}{\sqrt{4\pi\tau}} b_1\langle J\rangle \int_0^{s_0}
        \exp\left[ \frac{-(\hat{s}-t)^2}{4\tau}\right] \dif{t}
\\
       & + \frac{1}{\sqrt{4\pi\tau}} \exp\left( \frac{-\hat{s}^2}{4\tau}\right)
         \left[ c_0 \left\langle {\cal O}_6\right\rangle  - \frac{d_0 \hat{s}}{2\tau}
         \left\langle {\cal O}_8\right\rangle \right]
   \\
       &-\frac{16\pi^3}{\sqrt{4\pi\tau}} \int\dif{n}(\rho)\rho^4
       \int_0^{s_0} t^2  \exp\left[ \frac{-(\hat{s}-t)^2}{4\tau}\right]  J_2\left(\rho\sqrt{t} \right)
       Y_2\left(\rho\sqrt{t} \right)  \dif{t} \quad .
\end{split}
\end{equation}
The perturbative coefficients in these expressions are
\begin{gather}
 a_0 = -2\left(\frac{\alpha}{\ppi}\right)^2\left[1+\frac{659}{36}\frac{\alpha}{\ppi}+
  247.480\left( \frac{\alpha}{\ppi}\right)^2\right] \\
  a_1 = 2\left(\frac{\alpha}{\ppi}\right)^3\left[ \frac{9}{4}
            +65.781\frac{\alpha}{\ppi}\right] \quad ,\quad
  a_2 = -10.1250\left(\frac{\alpha}{\ppi}\right)^4
\end{gather}
as obtained from the three-loop  $\overline{\text{MS}}$ calculation of the correlation function \cite{che98}. 
As a result of renormalization group scaling of the GSRs \cite{orl00}, the coupling in  the perturbative coefficients is 
implicitly the running coupling at
the scale $\nu^2=\sqrt{\tau}$
 in the $\overline{\text{MS}}$ scheme
\begin{gather}
  \frac{\alpha (\nu^2)}{\pi} = \frac{1}{\beta_0 L}-\frac{\bar\beta_1\log L}{\left(\beta_0L\right)^2}+
  \frac{1}{\left(\beta_0 L\right)^3}\left[
  \bar\beta_1^2\left(\log^2 L-\log L -1\right) +\bar\beta_2\right]
  \label{run_coupling}\\
  L=\log\left(\frac{\nu^2}{\Lambda^2}\right)\quad ,\quad \bar\beta_i=\frac{\beta_i}{\beta_0}
  \quad ,\quad
  \beta_0=\frac{9}{4}\quad ,\quad \beta_1=4\quad ,\quad \beta_2=\frac{3863}{384}
\end{gather}
with
 $\Lambda_{\overline{MS}}\approx 300\,{\rm MeV}$ for three active flavours,
consistent with current estimates of $\alpha(M_\tau)$~\cite{pdg}.

 The condensate contributions in \bref{GM1}, \bref{G0}  involve 
next-to-leading order~\cite{bag90}
contributions\footnote{The calculation of next-to-leading contributions in \cite{bag90} have been extended non-trivially to 
$n_f=3$ from $n_f=0$, and the operator basis has been changed from $\left\langle \alpha G^2\right\rangle$ to $\langle J\rangle$.} from the dimension four
gluon condensate $\langle J\rangle$ and leading order~\cite{NSVZ_glue}
contributions from gluonic condensates of dimension six and eight
\begin{gather}
   \label{dimsix}
   \left\langle {\cal O}_6\right\rangle  =
    \left\langle g f_{abc}G^a_{\mu\nu}G^b_{\nu\rho}G^c_{\rho\mu}\right\rangle \\
   \left\langle {\cal O}_8\right\rangle = 14\left\langle\left(\alpha f_{abc}G^a_{\mu\rho}
   G^b_{\nu\rho}\right)^2\right\rangle
   -\left\langle\left(\alpha f_{abc}G^a_{\mu\nu}G^b_{\rho\lambda}\right)^2\right\rangle
   \label{dimeight}\\
 b_0 = 4\ppi\frac{\alpha}{\ppi}\left[ 1+ \frac{175}{36}\frac{\alpha}{\ppi}\right]
  \, ,\quad
  b_1 = -9\ppi\left(\frac{\alpha}{\ppi}\right)^2
  \, ,\quad
  c_0 = 8\ppi^2\left(\frac{\alpha}{\ppi}\right)^2
  \, ,\quad
  d_0 = 8\ppi^2\frac{\alpha}{\ppi} \quad.  \label{cond_coefficients}
\end{gather}
The remaining terms in the GSRs \bref{GM1}, \bref{G0} represent instanton contributions obtained from 
single instanton and anti-instanton~\cite{basic_instanton}
({\it i.e.,} assuming  that multi-instanton effects are negligible~\cite{schaefer_shuryak}) 
contributions to the scalar gluonic correlator~\cite{shuryak,gluelet,NSVZ_glue,inst_K2}.  
The quantity $\rho$ is the instanton radius,  $n(\rho)$
is the instanton density function, and $J_2$ and $Y_2$ are Bessel functions in the notation of \cite{abr}.

The instanton contributions to the Gaussian sum-rules can be interpreted \cite{gluelet} as naturally partitioning into an 
instanton continuum portion devolving from
\begin{equation}
\frac{1}{\pi}{\rm Im}\Pi^{inst}(t)=-16\pi^3\int \mathrm{d}n(\rho)\,\rho^4t^2
J_2\left(\rho\sqrt{t}\right) Y_2\left(\rho\sqrt{t}\right)
\end{equation}
and a contribution which, like the $\Pi(0)$ low-energy theorem (LET) term, appears only in the $k=-1$ sum-rule 
\begin{equation}
{
-\frac{128\pi^2}{\sqrt{4\pi\tau}} \exp\left( \frac{-\hat{s}^2}{4\tau}\right)
      \int\dif{n}(\rho)}\quad .
\label{ins_let}
\end{equation}  
This asymmetric role played by the instanton is crucial in obtaining a consistent analysis from these two sum-rules. 
In the absence of instanton contributions the LET tends to dominate the left-hand side of \bref{final_gauss}, corresponding to a 
massless state in the single-narrow resonance model \bref{phenom_single}.  However,  the higher-weighted sum-rules are independent of the LET, and 
in the absence of instantons lead to a much larger mass scale in  sum-rule analyses.  This discrepancy between the LET-sensitive and LET-insensitive
sum-rules has been shown to be resolved when instanton contributions are included in the Laplace sum-rules \cite{shuryak,gluelet}.   
A similar  qualitative behaviour emerges from the Gaussian sum-rules, since the LET term in \bref{final_gauss} and the LET-like instanton contribution
\bref{ins_let} have the same functional form [{\it i.e.,} each is proportional to $\exp\left( -\hat{s}^2/(4\tau)\right)$] and occur with opposite sign 
in the left-hand side of  \bref{final_gauss}.  Thus there exists a cancellation between these effects, which is easily verified as being significant
in the instanton liquid model~\cite{DIL} 
\begin{gather}
   n(\rho) = n_{\text{c}} \delta(\rho-\rho_{{\rm c}})
\\
 n_{\text{c}} = 8.0\times 10^{-4}\ \unt{GeV}^4 \quad ,\quad\ \rho_{{\rm c}} =
  \frac{1}{0.6}\ \unt{GeV}^{-1}\quad,
\end{gather}
along with a standard value for the gluon condensate ~\cite{nar97}
\begin{equation}\label{dimfour}
   \langle\alpha G^2\rangle
      = (0.07\pm 0.01)\ \unt{GeV}^4 \quad.
\end{equation}
This qualitative argument is upheld by the detailed GSR analysis presented in \cite{har01}.  However, the $k=-1$  GSR analysis is more sensitive to QCD uncertainties,  justifying a focus on the $k=0$ GSR for the remainder 
of this paper.  

\section{Gaussian Sum-Rule Analysis of Scalar Gluonium}
\begin{figure}[htb]
\centering
\includegraphics[scale=0.4]{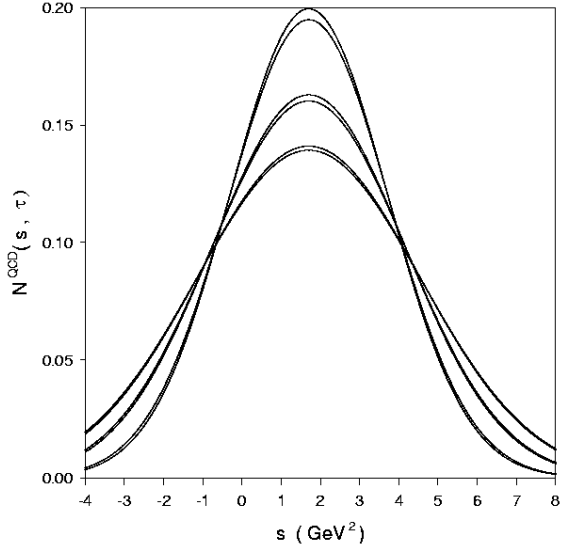}
\caption{ 
{ \small Comparison of the
theoretical prediction for $N_0^{QCD}\left(\hat s, \tau,s_0\right)$ with the
single narrow resonance phenomenological mode
The $\tau$ values used for the three pairs of curves, from top to bottom in the figure, are respectively
 $\tau=2.0\ \unt{GeV}^4$, $\tau=3.0\ \unt{GeV}^4$, and $\tau=4.0\ \unt{GeV}^4$.  Note the prominent disagreement between the QCD  
prediction and 
phenomenological model in the vicinity of the peaks.
}}
\label{sum_rules_single}
\end{figure}

\begin{figure}[htb]
\centering
\includegraphics[scale=0.35]{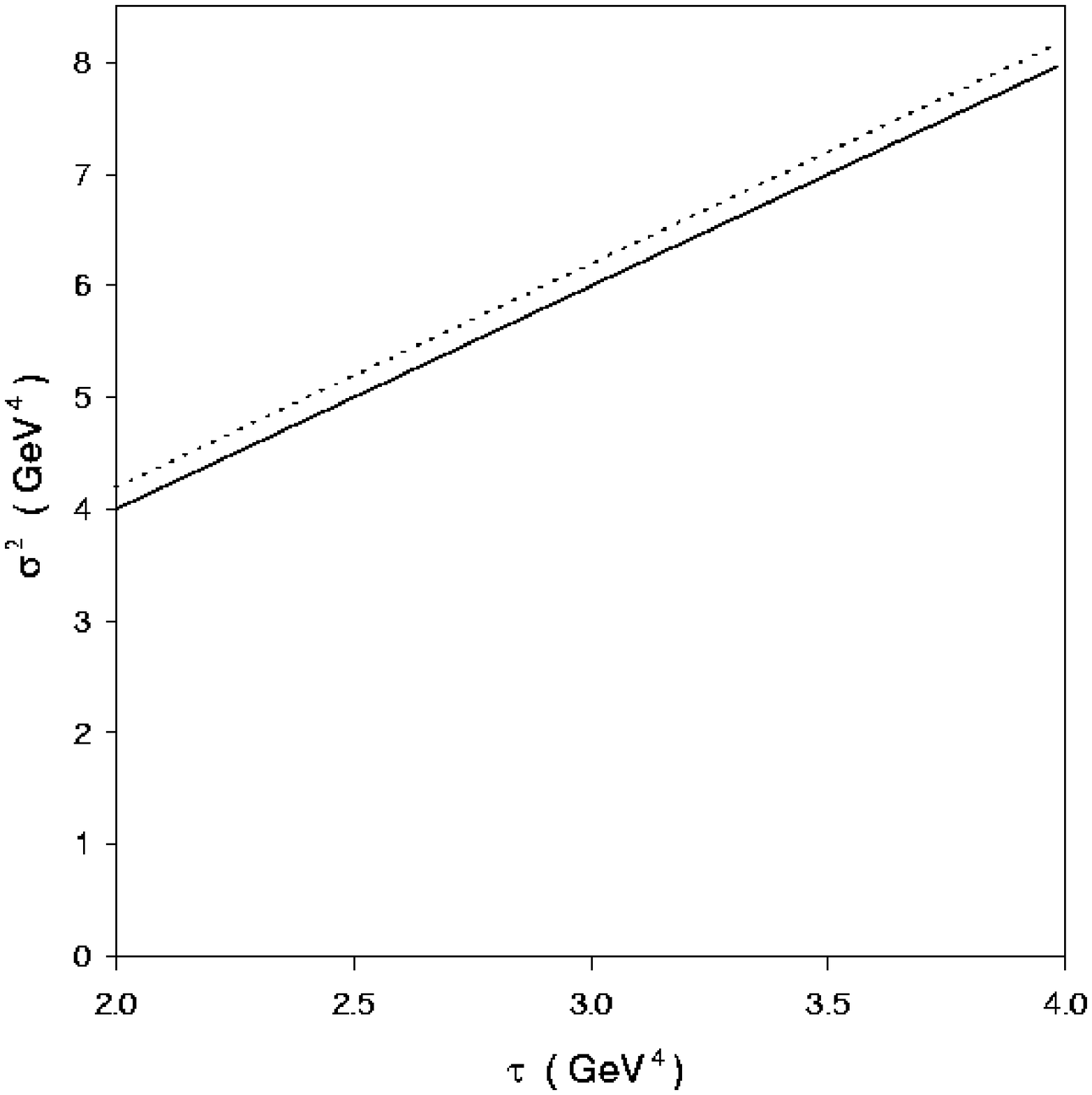}
\caption{
{ \small Plot of $\sigma_0^2$ for the theoretical prediction (dotted curve) compared with $\sigma_0^2=2\tau$
for the single-resonance model (solid curve) for the $k=0$ sum-rule.
}}
\label{sigma}
\end{figure}

 A single narrow resonance model analysis of the $k=0$ Gaussian sum-rule results in a mass scale of approximately $1.3\,{\rm GeV}$, but leads to poor   agreement
between the phenomenological model and QCD prediction as shown in Figure \ref{sum_rules_single}, indicating that the gluonium spectral function is poorly described by a single narrow 
resonance.
Furthermore, in the narrow width model the second-order moment combination \bref{square_params_b} should satisfy
\begin{equation}
\sigma_0^2-2\tau=0\quad, 
\end{equation}
but  Figure \ref{sigma} shows a substantial deviation from  this behaviour \cite{har01}.  Since this moment combination is related to the width of the GSR, we 
conclude that  the gluonium resonance strength must be distributed over a significant energy range. 

 The clear failure of the single narrow resonance 
model, indicative of distributed resonance strength significant enough to be resolved by the GSRs, is a significant conclusion in its own right, but
various distributed resonance strength models have also been analyzed \cite{har01}.  In order of increasing number of parameters (and increasing complexity)
they are
\begin{enumerate}
\item {Single non-zero width models} (2 parameters: mass, width)

\item {Two narrow resonance model} (3 parameters: two masses, relative resonance strength)

\item  {Narrow resonance plus a non-zero width resonance models} 
(4 parameters: two masses, one width, relative resonance strength)

\end{enumerate}

In the single non-zero width models, elaborations on the square pulse include a Gaussian resonance and a skewed Gaussian resonance models 
\begin{gather}
\rho(t)\sim \exp{\left[-\frac{\left(t-m^2\right)^2}{2\Gamma^2}\right]}
\label{gauss_res}
\\
\rho(t)\sim t^2\exp{\left[-\frac{\left(t-m^2\right)^2}{2\Gamma^2}\right]}
\label{skew_gauss_res}
\end{gather}
which are analytically and numerically simpler to analyze than a Breit-Wigner shape.  In the Gaussian resonance models, the quantity $\Gamma$ can be related to an equivalent Breit-Wigner width by  $\Gamma_{BW}=\sqrt{2\log 2}\Gamma/m$, and the  $t^2$ factor in the skewed Gaussian is chosen for consistency with (low-energy) two-pion decay rates \cite{let,pion}.  The relevant moment combinations for the Gaussian model \bref{gauss_res} are \cite{har01}
\begin{gather}
  \frac{M_{0,1}}{M_{0,0}} =m^2+\Gamma\Delta\label{gauss_m1} \\
  \sigma_0^2-2\tau=\Gamma^2-m^2\Gamma\Delta-\Gamma^2\Delta^2 \label{gauss_sigma} \\
  A^{(3)}_0=\left(m^4\Gamma-\Gamma^3\right)\Delta+3m^2\Gamma^2\Delta^2+2\Gamma^3\Delta^3 \label{gauss_A}
\end{gather}
where 
\begin{equation}\label{deldef}
  \Delta=\sqrt{\frac{2}{\pi}}\left[
  \frac{\exp{\left(-\frac{m^4}{2\Gamma^2}\right)}}{1+{\rm erf}
  \left(\frac{m^2}{\sqrt{2}\Gamma} \right)}\right]\quad.
\end{equation}
The quantity $\Delta$ is small, so the Gaussian model cannot accommodate large values of the asymmetry  $ A^{(3)}_0$.  However, the skewed Gaussian naturally leads to a larger asymmetry as reflected in the following results for the relevant moment combinations \cite{har01}
\begin{gather}
  \frac{M_{0,1}}{M_{0,0}} = \frac{m^2(m^4+3\Gamma^2)}{m^4+\Gamma^2}
  + \mathcal{O}(\Delta) \label{skew_m1}\\
  \sigma_0^2-2\tau = \frac{\Gamma^2(m^8+3\Gamma^4)}{(m^4+\Gamma^2)^2}
  + \mathcal{O}(\Delta) \label{skew_sigma}\\
  A_0^{(3)} = \frac{4m^2\Gamma^6(m^4-3\Gamma^2)}{(m^4+\Gamma^2)^3}
  + \mathcal{O}(\Delta) \quad .\label{skew_A}
\end{gather}

The results for the single non-zero resonance models are  shown in Table \ref{width_tab}, and indicate that the 
non-zero width models underestimate the QCD value of the asymmetry $ A_0^{(3)}$ by at least an order of magnitude \cite{har01}.  
This failure suggests that further phenomenological models which can generate a larger asymmetry are required.

\begin{table}[htb]
  \centering
  \begin{tabular}{||c|c|c|r||}
    \hline\hline
     & mass (GeV) & width (GeV) & $A^{(3)}_0~({\rm GeV^6})$ \\ 
    \hline\hline
  square pulse & $1.30\pm 0.17$ & $0.59\pm 0.07$ & 0 \\\hline  
  unskewed gaussian & $1.30\pm 0.17$ & $0.40\pm 0.05$ & $0.000342$ \\\hline
    skewed gaussian & $1.17\pm 0.15$ & $0.49\pm 0.06$ & {0.00943} \\\hline
QCD  & --- & ---& {-0.0825}\\
    \hline\hline
  \end{tabular}
 \caption{
{\small The results of a $k=0$ Gaussian sum-rules analysis
           of a variety of non-zero  resonance width models.  The  quoted resonance parameters include uncertainties introduced
by the QCD input parameters, except for $A^{(3)}_0$ which is obtained from the central values.
    For the Gaussian resonance models,
    the given width is  the equivalent Breit-Wigner width 
 }  }
\label{width_tab}
\end{table}

A model containing two narrow resonances 
\begin{equation}\label{double}
  \rho^{{\rm had}}(t) = \pi 
  \left[f_1^2 \delta(t-m_1^2) + f_2^2 \delta(t-m_2^2) \right]
\end{equation}
results in the following normalized Gaussian sum-rule~(\ref{ngsr})  
\begin{equation} \label{phenom_double}
  N^{\qcd}_0(\hat{s},\tau,s_0)  = \frac{1}{\sqrt{4\ppi\tau}}
  \left\{ r_1 \exp\left[-\frac{(\hat{s}-m_1^2)^2}{4\tau}\right]
  +       r_2 \exp\left[-\frac{(\hat{s}-m_2^2)^2}{4\tau}\right]
  \right\}\quad.
\end{equation}
where
\begin{equation}\label{crud}
  r_1 =  \frac{f_1^2 }{f_1^2 +f_2^2 } \quad,\quad
  r_2 =   \frac{f_2^2 }{f_1^2 +f_2^2 }\quad\,\quad
  r_1+r_2=1
\end{equation}
parameterize the relative strength of the two resonances.  This model can accommodate a large asymmetry parameter through an asymmetric distribution of resonance strength, as indicated by the following moment combinations \cite{har01}
\begin{gather}
  \frac{M_{0,1}}{M_{0,0}} = \frac{1}{2}(z+ry)\label{double_m1}\\
  \sigma_0^2-2\tau = \frac{1}{4}y^2(1-r^2) \label{double_sigma}\\
  A_0^{(3)} = -\frac{1}{4}ry^3(1-r^2)\label{double_A}\\
  S_0-12\tau^2 -12\tau\left(\sigma_0^2 -2\tau\right) 
  =\frac{1}{16}y^4\left(1-r^2\right)\left(1+3 r^2\right)\label{double_s} \quad,
\end{gather}
where
\begin{gather}\label{new_params}
  z  =  m_1^2 + m_2^2 \quad,\quad
  y  =  m_1^2 - m_2^2 \quad,\quad
  r  =  r_1 - r_2
\\
\label{skmom}
  S_0\equiv \frac{M_{0,4}}{M_{0,0}}-4\frac{M_{0,3}}{M_{0,0}}\frac{M_{0,1}}{M_{0,0}}
  +6\frac{M_{0,2}}{M_{0,0}}\left(\frac{M_{0,1}}{M_{0,0}}\right)^2-3
  \left( \frac{M_{0,1}}{M_{0,0}}\right)^4\quad.
\end{gather}
The lowest three moment combinations \bref{double_m1}--\bref{double_A} are used to determine the three resonance parameters, and the final fourth-order
residual combination \bref{double_s} is used as a test of the agreement between QCD and the phenomenological model.  The resulting 
resonance parameters are \cite{har01}
\begin{equation}
  m_1=(0.98\pm 0.2)\unt{GeV},~ m_2=(1.4\pm 0.2)\unt{GeV},~ r_1=0.28\mp 0.06,~ r_2=1-r_1 .
\label{two_res_numbers}
\end{equation}
Results from the fourth-order moment measure of the accuracy between QCD and the two-narrow resonance model is given in Table \ref{doubres_tab}, and indicate an approximately 50\% discrepancy between QCD and the phenomenological model \cite{har01}.

\begin{table}[htb]
  \centering
 
  \begin{tabular}{||c|c||}
    \hline\hline
     &   $S_0-12\tau^2 -12\tau\left(\sigma_0^2 -2\tau\right)$    \\ 
    \hline\hline
QCD    & $0.170\unt{GeV^8}$ \\\hline
double resonance model  & $0.074\unt{GeV^8}$\\
    \hline\hline
  \end{tabular}
 \caption{
{\small The results of a $k=0$ Gaussian sum-rules analysis
           of the double narrow resonance model
using central values of the QCD parameters.
 }  }
\label{doubres_tab}
\end{table}

As a  final attempt to improve the agreement between the next-highest moment test of the agreement between QCD and the phenomenological 
model, consider a narrow resonance of mass $m$ and relative strength $r_m$, combined with a second resonance of mass $M$, width $\Gamma$ and 
relative strength $r_M$.  Of course for a normalized GSR we must have $r_m+r_M=1$, so this defines a four-parameter model.
For example, when a square pulse is used to describe the second resonance the resulting normalized GSR is
\begin{equation}
\begin{split}
 N^{\qcd}_0(\hat{s},\tau,s_0)
=&r_m
\frac{1}{\sqrt{4\ppi\tau}}
 \exp\left[-\frac{(\hat{s}-m^2)^2}{4\tau}\right]
\\
&+r_M\frac{1}{4M\Gamma}\left[
{\rm erf}\left(\frac{\hat s-M^2+M\Gamma}{2\sqrt{\tau}}\right)
-{\rm erf}\left(\frac{\hat s-M^2-M\Gamma}{2\sqrt{\tau}}
\right)
\right]\quad ,
\end{split}
\label{N_narrow_square}
\end{equation}
where  $r_m+r_M=1$.  Four moment combinations are then needed to define the resonance parameters, and a fifth-order moment combination 
serves as a measure of the agreement between QCD and the phenomenological model \cite{har01}
\begin{gather}
 \frac{M_{0,1}}{M_{0,0}}=\frac{1}{2}\left( z+ ry \right)+ {\cal O}\left(\Delta\right)
\label{narrow_square_m1}
\\
\sigma_0^2-2\tau=\frac{1}{4}y^2\left(1-r^2\right)+\frac{1}{12}\Gamma^2\left(z-y\right)\left(1-r\right)
+ {\cal O}\left(\Delta\right)
\label{narrow_square_m2}
\\
A_0^{(3)}=-\frac{1}{4}y^3r\left(1-r^2\right)-\frac{1}{8}\Gamma^2
y\left(1-r^2\right)\left(z-y\right) + {\cal O}\left(\Delta\right)
\label{narrow_square_m3}
\\
\begin{split}
S_0-12\tau^2-12\tau\left(\sigma_0^2-2\tau\right)=&\frac{1}{16}y^4\left(1-r^2\right)\left(1+3r^2\right)
+\frac{1}{8}\Gamma^2y^2\left(1+r\right)\left(1-r^2\right)\left(z-y\right)
\\&
+\frac{1}{40}\Gamma^4\left(1-r\right)\left(z-y\right)^2+ {\cal
  O}\left(\Delta\right)
\end{split}
\label{narrow_square_m4}
\\
\begin{split}
A_0^{(5)}-20\tau A_0^{(3)}=&-\frac{1}{8}y^5r\left(1-r^2\right)\left(1+r^2\right)
-\frac{5}{48}\Gamma^2y^3\left(1-r^2\right)\left(1+r\right)^2\left(z-y\right)
\\
&-\frac{1}{16}\Gamma^4y\left(1-r^2\right)\left(z-y\right)^2+ {\cal O}\left(\Delta\right)
\end{split}
\label{narrow_square_m5}
\end{gather} 
where the fifth-order asymmetry moment is
\begin{equation}
A_k^{(5)}= \frac{M_{k,5}}{M_{k,0}}-5 \frac{M_{k,4}}{M_{k,0}}
\frac{M_{k,1}}{M_{k,0}}
+10 \frac{M_{k,3}}{M_{k,0}}\left( \frac{M_{k,1}}{M_{k,0}}\right)^2
-10 \frac{M_{k,2}}{M_{k,0}}\left( \frac{M_{k,1}}{M_{k,0}}\right)^3
+4\left( \frac{M_{k,1}}{M_{k,0}} \right)^5 
\quad .
\label{fifth_moment}
\end{equation}
The resulting resonance parameters and the QCD prediction and phenomenological values of the fifth-order asymmetry parameter are shown in Table \ref{res_summary_table} \cite{har01}.  An interesting feature of the results are that the wide  resonance is consistently found to be lighter than the narrow resonance, and that the narrow plus skewed Gaussian model shows only a 20\% deviation from the QCD value of the fifth order 
asymmetry parameter. 

{
\begin{table}
\centering
\begin{tabular}{||c|c|c|c|c|c||}
\hline\hline
 & $m\,({\rm GeV})$ &  $M\,({\rm GeV})$ & Width (GeV) & $r_m$ & $A_0^{(5)}-20\tau A_0^{(3)}$\\\hline\hline
 square & $1.33\pm 0.18$ & $1.23\pm 0.18$ & $0.95\pm 0.12$ & $0.6\pm 0.13$ & $-0.11\unt{GeV^{10}}$\\\hline
gauss & $1.41\pm 0.19$ & $1.23\pm 0.15$ & $0.52\pm 0.06$ & $0.49\pm 0.13$ & $-0.13\unt{GeV^{10}}$\\\hline
skewed  & $1.38\pm 0.13$ & $1.06\pm 0.21$ & $0.69\pm 0.07$ & $0.44\pm 0.04$ &${-0.19\unt{GeV^{10}}}$ \\\hline
QCD  &--- &--- &--- & --- & ${-0.24\unt{GeV^{10}}}$ \\\hline
\hline   
\end{tabular}
\caption{{ \small 
Resonance parameters obtained  
in the various two-resonance scenarios.
The label ``square'' denotes the narrow plus square pulse model, ``gauss'' refers to narrow plus Gaussian resonance model, and ``skewed'' indicates the narrow plus skewed Gaussian model.
 The mass $M$ denotes the state
associated with the quoted width, which  corresponds  to the equivalent Breit-Wigner width for the Gaussian models.
The  quoted resonance parameters include uncertainties introduced
by the QCD input parameters, except for the fifth-order residual moment combination which is obtained from the central values.
}}
\label{res_summary_table}
\end{table}
}

\section{Comparison of Distributed Resonance Strength Models}
\begin{figure}[htb]
\centering
\includegraphics[scale=0.4]{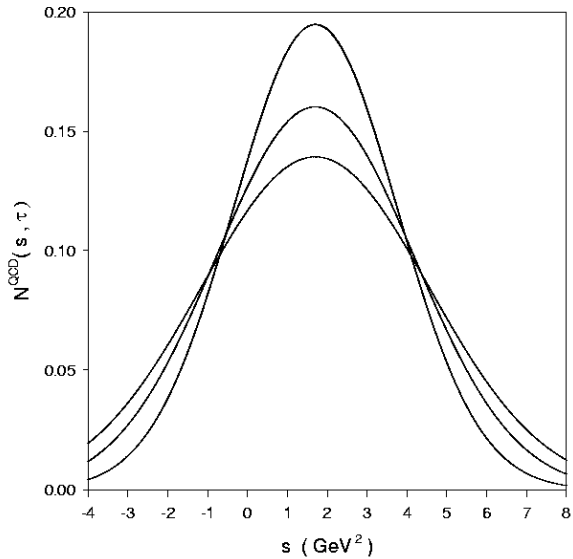}
\caption{
{  \small Comparison of the
theoretical prediction for $N_0^{QCD}\left(\hat s, \tau,s_0\right)$ with the
distributed resonance phenomenological models.
The $\tau$ values used for the three pairs of curves, from top to bottom in the figure, are respectively
 $\tau=2.0\ \unt{GeV}^4$, $\tau=3.0\ \unt{GeV}^4$, and $\tau=4.0\ \unt{GeV}^4$.
The almost perfect overlap between QCD and phenomenology is particularly impressive because there are no free parameters corresponding to the normalization of the QCD prediction and phenomenological models.
}}
\label{sum_rules_dist}
\end{figure}

All the distributed resonance strength models considered in the previous section lead to the excellent agreement between the 
QCD and phenomenological sides of the normalized GSR shown in Figure \ref{sum_rules_dist}, and clearly resolve the discrepancy evident in 
  Figure \ref{sum_rules_single} corresponding to the single resonance model \cite{har01}.   However, a $\chi^2$ measure of the agreement between the theoretical and phenomenological curves in  
Figure \ref{sum_rules_dist} result in a $\chi^2$ which is an order of magnitude smaller in the two-resonance models \cite{har01}.  
The combination of this result with the Table \ref{width_tab} observation of 
at least an order of magnitude  disagreement with the QCD value of the third-order asymmetry is compelling evidence in favour of a 
two-resonance scenario for distributed resonance strength, with an upper bound on the heavier mass of about $1.6\unt{GeV}$ obtained from
Eq.\ \bref{two_res_numbers} and Table \ref{res_summary_table}.

The various two-resonance models all have  similar values of 
$\chi^2$, which does not provide a useful criteria to distinguish between these models.  However, the narrow plus skewed Gaussian resonance model leads to  the best ($\approx 20\%$)  agreement with the 
QCD value of the next-order moment. In this scenario, the mass predictions and the pattern of a lighter broad resonance and a heavier narrow resonance is consistent with the identification of gluonium content in the $f_0(1370)$ and $f_0(1500)$.

\section{Conclusions}
In this paper, the formulation of Gaussian sum-rules has been reviewed.  The key qualitative feature of Gaussian sum-rules is their enhanced sensitivity to excited states in comparison with Laplace sum-rules.  Thus any resonance strong enough to stand out from the QCD continuum 
should reveal itself in a GSR analysis.  The significance of normalized GSRs has been emphasized since they provide information independent
of the the finite-energy sum-rule constraint that arises from the evolution of GSRs through the diffusion equation~\cite{orl00}.

Methods for obtaining phenomenological predictions from GSRs have been reviewed and applied to the $\rho$ meson to demonstrate the predictive power of GSRs  \cite{orl00}.  These techniques are easily adapted to a variety of resonance models, and through combinations of GSR moments, provide a natural criterion for testing the accuracy of the phenomenological model in comparison with QCD \cite{har01}.

The important role of instanton effects in relation to the low-energy theorem  has been illustrated for the GSRs of scalar gluonic 
currents.  This analysis provides additional support for the  interpretation of instanton effects first developed for Laplace 
sum-rules \cite{gluelet}. An identical, and natural, partitioning of instanton effects into a continuum and LET-like contribution resolves discrepancies between the LET-sensitive and LET-insensitive sum-rules in both the Laplace and Gaussian sum-rules.

A detailed phenomenological analysis of the $k=0$ GSR is presented because it is is less sensitive to QCD parameter uncertainties than the 
$k=-1$  (LET-dependent) GSR.  The analysis clearly rules out a single narrow resonance scenario, and further indicates that gluonium resonance strength is spread over a broad energy region.    In particular, evidence exists for two resonances below $\approx 1.6\unt{GeV}$ with a significant gluonium content, with the lighter resonance having a substantial width \cite{har01}.  Such a scenario is consistent gluonium content  of
$f_0(1370)$ and $f_0(1500)$.

\vskip 1cm
{\noindent \large\bf Acknowledgments:}\\
TGS and DH are grateful for research funding from the Natural Sciences \& Engineering Research Council of Canada (NSERC).
Many thanks to Amir Fariborz and SUNY IT for careful organization and generous  hospitality during the workshop on Theoretical High-Energy Physics.


\begin{thebibliography}{10}

\bibitem{pdg} K. Hagiwara {\it et al.}, Phys.\ Rev.\ \tref{D66}{2002}{010001}.

\bibitem{gauss} R.A.\ Bertlmann, G.\ Launer, E.\ de Rafael, Nucl.\ Phys.\ \tref{B250}{1985}{61}. 

\bibitem{orl00}
G. Orlandini, T.G. Steele and D. Harnett, Nucl.~Phys.\ \tref{A686}{261}{2001}.

\bibitem{har01} D.\ Harnett, T.G.\ Steele, Nucl.\ Phys.\  \tref{A695}{2001}{205}.

\bibitem{shuryak} E.V.\ Shuryak, Nucl.\ Phys.\ \tref{B203}{1982}{116};\\
H.\ Forkel, Phys.\ Rev.\ \tref{D64}{2001}{034015}.


\bibitem{gluelet} D.\ Harnett, T.G.\ Steele, V.\ Elias, Nucl.\ Phys.\ \tref{A686}{2001}{393}.

\bibitem{let}
V.A. Novikov, M.A. Shifman, A.I. Vainshtein and V.I. Zakharov, Nucl.~ Phys. \tref{B191}{1981}{301}.
 
\bibitem{dim_eight}   V.\ Gim\'enez, J.\ Bordes, J.\ Pe\~narrocha, Phys.\ Lett.\ \tref{B223}{1989}{251};\\
S.N.\ Nikolaev, H.R.\ Radyushkin, Nucl.\ Phys.\ \tref{B213}{1983}{285};\\
D.J.\ Broadhurst, S.C.\ Generalis, Phys.\ Lett.\ \tref{B165}{1985}{175}.

\bibitem{bordes}  V.\ Gim\'enez, J.\ Bordes, J.\ Pe\~narrocha, Nucl.\ Phys.\ \tref{B357}{1991}{3}. 


\bibitem{che98}
K.G. Chetyrkin, B.A. Kneihl and M. Steinhauser, Nucl.~Phys.\ \tref{B510}{1998}{61}.

\bibitem{basic_instanton}
A.\ Belavin, A.\ Polyakov, A.\ Schwartz and Y.\ Tyupkin,
   Phys.\ Lett.\ \tref{B59}{1975}{85};\\
G.\ 't Hooft, Phys.\ Rev.\ \tref{D14}{1976}{3432}.

\bibitem{bag90}
E. Bagan and T.G. Steele, Phys.\ Lett.\ \tref{B234}{1990}{135}.

\bibitem{NSVZ_glue}
V.A.\ Novikov, M.A.\ Shifman, A.I.\ Vainshtein and V.I.\ Zakharov,
      Nucl.\ Phys.\ \tref{B165}{1980}{67}.

\bibitem{schaefer_shuryak}
T. Sch\"{a}efer and E.V. Shuryak, Phys.~Rev.~Lett.\ \tref{75}{1995}{1707}.


\bibitem{inst_K2}
B.V. Geshkenbein and B.L. Ioffe, Nucl.~Phys.\ \tref{B166}{1980}{340};\\
B.L. Ioffe and A.V. Samsonov, Phys.~of~Atom.~Nucl.\ \tref{63}{2000}{1527}.  


\bibitem{abr}
M. Abramowitz and I.E. Stegun, {\em Mathematical Functions with Formulas,
   Graphs, and Mathematical Tables} (National Bureau of Standards
   Applied Mathematics Series, Washington) 1972.


\bibitem{DIL}
E.V. Shuryak, Nucl.~Phys.\ \tref{B203}{1982}{93}.


\bibitem{nar97}
S. Narison, Nucl.~Phys.~B~(Proc.~Supp.) \tref{54A}{1997}{238}.


\bibitem{pion} V.A. Novikov, M.A. Shifman, Z. Phys. \tref{C8}{1981}{43};\\
M.A. Shifman, Z. Phys. \tref{C9}{1981}{347}.


\end{thebibliography}
\end{document}